\documentclass[a4paper]{jpconf}

\PassOptionsToPackage{usenames,dvipsnames}{xcolor}

\usepackage{graphicx}
\usepackage{citesort}
\usepackage{cite}
\usepackage{hyperref}
\hypersetup{colorlinks, linkcolor={blue}, citecolor={blue}, urlcolor={blue}}
\usepackage{xcolor}
\usepackage{multicol}
\usepackage{multirow}

\usepackage[utf8]{inputenc}
\usepackage[T1]{fontenc}
\usepackage{amsmath, amsfonts, amssymb}
\usepackage{mathtools}
\usepackage{booktabs}
\usepackage{pifont}

\newcommand{\cmark}{\ding{51}}
\newcommand{\xmark}{\ding{55}}

\DeclareUnicodeCharacter{27E8}{\ensuremath{\langle}}
\DeclareUnicodeCharacter{27E9}{\ensuremath{\rangle}}
\DeclareUnicodeCharacter{0394}{\ensuremath{\Delta}}
\DeclareUnicodeCharacter{00B2}{\ensuremath{^2}}

\usepackage[frozencache, cachedir=./]{minted}  
\makeatletter
\usepackage{etoolbox}
\AtBeginEnvironment{Verbatim}{
   \def\PYGvs@tok@err {\color{black} \def\PYGvs@bc##1{\strut ##1}}
}
\makeatother

\AtBeginEnvironment{minted}{%
  \catcode`?\active
  \begingroup\lccode`~=`\?\lowercase{\endgroup\def~{\linebreak}}%
}

\begin{document}

\title{\textsc{Lips}: $\boldsymbol{p}\kern0.4mm$-adic and singular phase space}

\author{Giuseppe De Laurentis}

\address{Paul Scherrer Institut, CH-5232 Villigen PSI, Switzerland}

\ead{giuseppe.de-laurentis@psi.ch\vspace{-2mm}}

\begin{abstract}
I present new features of the open-source Python package
\texttt{lips}, which leverages the newly developed \texttt{pyadic} and
\texttt{syngular} libraries. These developments enable the generation
and manipulation of massless phase-space configurations beyond real
kinematics, defined in terms of four-momenta or Weyl spinors, not only
over complex numbers ($\mathbb{C}$), but now also over finite fields
($\mathbb{F}_p$) and $p\kern0.2mm$-adic numbers ($\mathbb{Q}_p$).  The
package also offers tools to evaluate arbitrary spinor-helicity
expressions in any of these fields. Furthermore, using the
algebraic-geometry submodule, which utilizes \texttt{Singular}
\cite{DGPS} through the Python interface \texttt{syngular}, one can
define and manipulate ideals in spinor variables, enabling the
identification of irreducible surfaces where scattering amplitudes
have well-defined zeros and poles. As an example application, I
demonstrate how to infer valid partial-fraction decompositions from
numerical evaluations.
\end{abstract}
\vspace{-8mm}

\section{Introduction}
High-multiplicity loop-level amplitude computations involve
significant algebraic complexity, which is usually side-stepped via
numerical routines. Yet, when available, compact analytical
expressions can display improved numerical stability and reduced
evaluation times. Moreover, much of the recent progress in the
computation of loop-corrections to scattering amplitudes has been
achieved thanks to finite-field methods \cite{vonManteuffel:2014ixa,
  Peraro:2016wsq}. As these numerical computations are unsuited for
direct phenomenological applications, analytic expressions must be
recovered, so that they can then be evaluated with floating-point
numbers.

An important role to manifest the analytic properties of scattering
amplitudes is played by the spinor-helicity formalism. A classical
example is that numerators in gauge-theoretic amplitudes, such as
quantum chromodynamics, mitigate the degree of divergences which would
naively be expected from Feynman propagators, namely from factors of
$1/s_{ij}$ to $1/\sqrt{s_{ij}}$. Relaxing the constraint of real
kinematics, one realizes that these divergences are in fact either
purely holomorphic spinor contractions $\langle ij \rangle$ or purely
anti-holomorphic ones $[ij]$ (with $s_{ij} = \langle ij \rangle[ji]$).

It has been shown that significant insights into the analytic
structure of amplitudes can be obtained from tailored numerical
evaluations in singular limits \cite{DeLaurentis:2019bjh,
  DeLaurentis:2022otd}. For example, the rational prefactors appearing
in the planar two-loop amplitude for the process $0 \rightarrow
q^+\bar q^-\gamma^+\gamma^+\gamma^+$ with a closed fermion loop
\cite{Abreu:2020cwb, Chawdhry:2020for} can be simplified to just the
following two functions
\begin{align}
  & \frac{⟨23⟩[23]⟨24⟩[34]}{⟨15⟩⟨34⟩⟨45⟩⟨4|1+5|4]}+(45
  \rightarrow 54), \label{eq:rational-functions-1} \\
  & \frac{⟨13⟩[13]⟨24⟩[45]}{⟨13⟩⟨34⟩⟨45⟩⟨4|1+3|4]}+(45
  \rightarrow 54) - \frac{⟨12⟩[13]⟨23⟩^2}{⟨13⟩⟨24⟩⟨25⟩⟨34⟩⟨35⟩},
\end{align}
together with those obtained by closing the vector space generated by
these two functions under the permutations of the photons (legs 3, 4,
and 5). In a soon-to-appear paper, we obtain analogous expressions for
the full-color two-loop $0 \rightarrow q\bar q\gamma\gamma\gamma$
amplitude \cite{Abreu:2023xxx}.

\pagebreak

\section{\textsc{Lips}: a phase-space generator for theory computations}\label{sec:geometry}
Phase-space generators in high-energy physics traditionally describe
the kinematics of physical processes at colliders, meaning they
provide real-valued phase-space configurations. Yet, from a
theoretical standpoint, the analytic properties of scattering
amplitudes in perturbative quantum field theory are better understood
in the complex plane. This motivates the development of a phase-space
generator that exploits the additional freedom of complex kinematics.

The package \texttt{lips} (short for Lorentz invariant phase space)
provides a phase-space generator and manipulator that is tailored to
the needs of modern theoretical calculations. The package is designed
to work for processes of arbitrary multiplicity, although at present
it can easily handle massless particles only. Nevertheless, massive
particles can already be described in terms of a pair of massless
decay products. Use cases include: 1) generation of phase-space points
over complex numbers ($\mathbb{C}$), finite fields ($\mathbb{F}_p$),
and $p\kern0.1mm$-adic numbers ($\mathbb{Q}_p$); 2) on-the-fly
evaluation of arbitrary spinor-helicity expressions; 3) construction
of special kinematic configurations; 4) algebro-geometric analysis of
irreducible varieties in kinematic space.

Live examples powered by \texttt{binder} \cite{binder} are accessible
through the badge on the GitHub page.\footnote{At the URL
  \href{https://github.com/GDeLaurentis/lips}{github.com/GDeLaurentis/lips}}

\subsection{Installation}
The required language is Python 3, with version $\geq 3.8$ being
recommended. The package is available through the Python Package
Index,\footnote{At the URL
  \href{https://pypi.org/project/lips/}{pypi.org/project/lips/}} and
thus can be installed via \texttt{pip}.
\begin{minted}[escapeinside=||, mathescape=true, linenos=False, numbersep=5pt, gobble=2, frame=lines, framesep=2mm, breaklines, breakautoindent=false, breakindent=-12.5pt, baselinestretch=0 ]{shell}
  pip install --upgrade lips
\end{minted}
\vspace{-1.5mm}
Alternatively, the source code can be cloned from GitHub,
and then installed with \texttt{pip}.
\begin{minted}[escapeinside=||, mathescape=true, linenos=False, numbersep=5pt, gobble=2, frame=lines, framesep=2mm, breaklines, breakautoindent=false, breakindent=-12.5pt, baselinestretch=0 ]{shell}
  pip install -e path/to/repository
\end{minted}
\vspace{-1.5mm}
The option \texttt{-e} ensures that changes to the source code are
reflected without having to reinstall the package, for instance after
invoking \texttt{git pull} to download an update.

\subsection{Dependencies}
The Python ecosystem provides a rich variety of third-party,
open-source libraries for scientific computing. Among these
\texttt{lips} depends on \texttt{NumPy} \cite{2020NumPy-Array}, whose
\texttt{ndarray} class is used to describe all Lorentz tensors,
\texttt{mpmath} \cite{mpmath}, from which multi-precision real and
complex numbers are imported, and \texttt{sympy}
\cite{10.7717/peerj-cs.103}, for symbolic manipulations. Dependencies
are declared in the file \texttt{setup.py} and are installed
automatically through \texttt{pip}. The only exception is
\texttt{Singular} \cite{DGPS}, which is optional, and needs to be
installed separately. Additionally two new, open-source dependencies
are used, namely \texttt{pyadic} and \texttt{syngular}. They can be
used independently of \texttt{lips}.

\subsubsection{\normalfont\texttt{pyadic}}
Finite fields have become a staple of multi-loop computations. More
recently, the idea to use $p\kern0.1mm$-adic numbers was introduced,
in order to rescue a non-trivial absolute value while maintaining
integer arithmetic \cite{DeLaurentis:2022otd}. The package
\texttt{pyadic} provides classes for finite fields (\texttt{ModP}) and
$p\kern0.1mm$-adic numbers (\texttt{PAdic}), instantiated as shown, as
well as related algorithms.
\begin{minted}[escapeinside=||, mathescape=true, linenos=False, numbersep=5pt, gobble=2, frame=lines, framesep=2mm, breaklines, breakautoindent=false, breakindent=-12.5pt, baselinestretch=0 ]{python}
  ModP(|\color{OliveGreen}{|number|}|, |\color{OliveGreen}{|prime|}|)
  PAdic(|\color{OliveGreen}{|number|}|, |\color{OliveGreen}{|prime|}|, |\color{OliveGreen}{|digits|}|, |\color{OliveGreen}{|valuation|}|=|\color{OliveGreen}{|0|}|)
\end{minted}
\vspace{-1.5mm} Besides standard arithmetic operations, square roots
are also available in the functions \texttt{finite\_field\_sqrt} and
\texttt{padic\_sqrt}. These are not guaranteed to be in the field and
may return a \texttt{FieldExtension} object. Only field extensions by
a single square root are currently implemented. Moreover, the
$p\kern0.1mm$-adic logarithm is also implemented in the function
\texttt{padic\_log}.

For $p\kern0.1mm$-adic numbers, numerical precision is explicitly tracked as an
error $\mathcal{O}(p^k)$ term, meaning all displayed digits are, by
default, significant digits. This allows one to perform computations
in singular configurations while keeping track of the numerical
uncertainty. Nevertheless, the parameter
\texttt{fixed\_relative\_precision} can be switched to \texttt{True}
to emulate the usual floating-point behavior, with numerical noise
being appended to numbers in case of precision loss.

Rational reconstruction algorithms from $\mathbb{F}_p$ and
$\mathbb{Q}_p$ to $\mathbb{Q}$ are also provided with the function
\texttt{rationalise}, which takes an optional keyword argument
\texttt{algorithm} to toggle between maximal quotient reconstruction
(\texttt{MQRR}) and lattice reduction (\texttt{LGRR})
\cite{Lenstra1982, inproceedings, Klappert:2019emp}.


It is also worth mentioning that an alternative implementation of
finite fields is available in the package \texttt{galois}
\cite{Hostetter_Galois_2020}. This is particularly useful when dealing
with large \texttt{ndarrays}.

\subsubsection{\normalfont\texttt{Singular}/\texttt{syngular}}
The submodule \texttt{algebraic\_geometry} of \texttt{lips} requires
\texttt{Singular}~\cite{DGPS}.\linebreak To facilitate computations,
an object-oriented Python interface is provided in the package
\texttt{syngular} \cite{syngular}. Like \texttt{pyadic}, this can be
used independently of \texttt{lips}.  The main classes currently
implemented are \texttt{Ideal}, \texttt{Ring} and
\texttt{QuotientRing}. They provide easy access to several functions
implemented in \texttt{Singular} in a pythonic way. For instance, we
have ideal addition (\texttt{+}), product (\texttt{*}), quotient
(\texttt{/}), membership (\texttt{in}), intersection (\texttt{\&}),
equality (\texttt{==}), \textit{etc}. More methods are available with
self-explanatory names, e.g.~\texttt{primary\_decomposition}, which
calls \texttt{primdecGTZ}.

\subsection{Basic usage}

The main class provided by \texttt{lips} is the \texttt{Particles}
class. It describes the phase space of massless particles with given
{\color{OliveGreen}\texttt{multiplicity}} in 4 dimensions. By default,
the 4-momenta are taken to be complex valued. A specific choice of
number field can be passed as a keyword parameter.
\begin{minted}[escapeinside=||, mathescape=true, linenos=False, numbersep=5pt, gobble=2, frame=lines, framesep=2mm, breaklines, breakautoindent=false, breakindent=-12.5pt, baselinestretch=0 ]{python}
  Particles(|\color{OliveGreen}{|multiplicity|}|, field=Field(|\color{BrickRed}{|name|}|, |\color{OliveGreen}{|prime|}|, |\color{OliveGreen}{|digits|}|))
\end{minted}
\vspace{-1.5mm}
The generated phase-space point will satisfy on-shell relations
($p^\mu_i p_{i,\mu} = 0 \; \forall \; i$) and momentum conservation
($\sum_i p_i^\mu =0$). Valid choices for the \texttt{field} keyword
are multi-precision complex numbers ($\mathbb{C}$), Gaussian rationals
($\mathbb{Q}[i]$), finite fields ($\mathbb{F}_p$), and $p\kern0.1mm$-adic
numbers ($\mathbb{Q}_p$).
\vspace{-4mm}
\begin{multicols}{2}
\begin{minted}[escapeinside=||, mathescape=true, linenos=False, numbersep=5pt, gobble=2, frame=lines, framesep=2mm, breaklines, breakautoindent=false, breakindent=-12.5pt, baselinestretch=0 ]{python}
  Field("mpc", |\color{OliveGreen}{|0|}|, |\color{OliveGreen}{|300|}|) |\\[0.5mm]|
  Field("gaussian rational", |\color{OliveGreen}{|0|}|, |\color{OliveGreen}{|0|}|)
\end{minted}
\columnbreak
\begin{minted}[escapeinside=||, mathescape=true, linenos=False, numbersep=5pt, gobble=2, frame=lines, framesep=2mm, breaklines, breakautoindent=false, breakindent=-12.5pt, baselinestretch=0 ]{python}
  Field("finite field", |\color{OliveGreen}{|2147483647|}|, |\color{OliveGreen}{|1|}|) |\\[0.5mm]|
  Field("padic", |\color{OliveGreen}{|2147483629|}|, |\color{OliveGreen}{|3|}|)
\end{minted}
\end{multicols}
\vspace{-2mm}
\noindent Finite fields and $p\kern0.1mm$-adic numbers are taken from
a specific slice of complex phase space, namely that with $E, p_x, p_z
\in \mathbb{Q}$ and $p_y \in i\mathbb{Q}$. This is equivalent to a
change of metric to $\text{diag}(1,-1,1,-1)$.  Depending on the choice
of field, some parameters are discarded (see
Table~\ref{Tab:FieldArgs}).

\begin{table}[t]
  \centering
  \begin{tabular}{ccccc}
    \toprule
    & {\color{BrickRed}\texttt{"mpc"}} & {\color{BrickRed}\texttt{"gaussian rational"}} & {\color{BrickRed}\texttt{"finite field"}} & {\color{BrickRed}\texttt{"padic"}}   \\
    \midrule
    {\color{OliveGreen}\texttt{prime}} & \xmark & \xmark & \cmark & \cmark \\
    {\color{OliveGreen}\texttt{digits}} & \cmark & \xmark & \xmark & \cmark \\
    \bottomrule
  \end{tabular}
  \caption{\label{Tab:FieldArgs}Used (\cmark) and discarded (\xmark)
    arguments for \texttt{Field}.}
\end{table}

\begin{table}[t]
  \centering
  \begin{tabular}{c|c|c || c|c|c}
    \toprule
    Lorentz repr. & symbol & \texttt{Particle} property &  Lorentz repr. & symbol. & \texttt{Particle} property  \\
    \midrule
    \multirow{2}{*}{$(0, 1/2)$} & $\lambda^\alpha$ & \texttt{r\_sp\_u} & \multirow{4}{*}{$(1/2, 1/2)$} & $p^{\mu}$ & \texttt{four\_mom} \\
                                                  & $\lambda_\alpha$ & \texttt{r\_sp\_d} & & $p_{\mu}$ & \texttt{four\_mom\_d}\\
    \cmidrule{1-3}\cmidrule{5-6}
    \multirow{2}{*}{$(1/2, 0)$} & $\bar\lambda^{\dot\alpha}$ & \texttt{l\_sp\_u} & & $p^{\dot\alpha\alpha}$ & \texttt{r2\_sp}\\
                                & $\bar\lambda_{\dot\alpha}$ & \texttt{l\_sp\_d} & & $\bar p_{\alpha\dot\alpha}$ & \texttt{r2\_sp\_b}\\
    \bottomrule
  \end{tabular}
  \caption{\label{Tab:LorentzGroupRepresenations}Representations of
    the Lorentz group and associated properties of the
    \texttt{Particle} class.}
\end{table}

The \texttt{Particles} class is a 1-indexed subclass of the Python
built-in type \texttt{list}, with \texttt{Particle} class
entries. While this re-indexing may be unusual in Python, it is the
natural choice to match the notation used to write scattering
amplitudes. The \texttt{Particle} objects have several attributes,
each corresponding to one of the representations of the Lorentz
group. For instance, we have right spinors with index up
(\texttt{r\_sp\_u}), left spinors with index down (\texttt{l\_sp\_d}),
rank-two spinors (\texttt{r2\_sp}), four momenta (\texttt{four\_mom}),
\textit{etc.} See Table~\ref{Tab:LorentzGroupRepresenations} for a
more schematic representation. Updating the \linebreak values for one
of these properties automatically updates the values of the rest. The
spinor conventions employed are fairly common in the field, for more
details please see ref.~\cite[Section
  2.2]{DeLaurentis:2020xar}.\linebreak
\indent Another basic functionality provided in \texttt{lips} is the
evaluation of arbitrary spinor-helicity expressions. This works
seamlessly for all of the above defined fields. Understood symbols
include arbitrary spinor strings, with limited support for open-index
expressions, Mandelstam variables with any numbers of indices, Gram
determinants of three-mass triangle diagrams
(\texttt{Δ\_ij|kl|mn}),\footnote{See K\"all\'en function or Heron's
  formula.}\linebreak and traces involving $\gamma^5$
(\texttt{tr5\_ijkl}). This feature can be accessed is via the
\texttt{\_\_call\_\_} magic method of the $\texttt{Particle}$ class,
as shown.
\begin{minted}[escapeinside=||, mathescape=true, linenos=False, numbersep=5pt, gobble=2, frame=lines, framesep=2mm, breaklines, breakautoindent=false, breakindent=-12.5pt, baselinestretch=0 ]{python}
  Particles(5)("(8/3s23⟨24⟩[34])/(⟨15⟩⟨34⟩⟨45⟩⟨4|1+5|4])")
\end{minted}
\vspace{-1.5mm}
Regular expressions (with the package \texttt{re}) are used to split
the string into individual invariants, which then form an abstract
syntactic tree (with the package \texttt{ast}). The individual
invariants are computed in the \texttt{Particles.compute} method. For
simplicity's sake, greater-than and less-than symbols can be used in
lieu of the angle brackets to denote holomorphic spinor contractions.

Another useful method is the \texttt{Particles.image} method, which
implements transformations under the symmetries of phase space. These
are represented by a tuple whose first entry is a string representing
a permutation of the external legs, followed by a Boolean denoting
whether a swap of the left and right representations of the Lorentz
group is needed ($\lambda_\alpha \leftrightarrow
\bar\lambda_{\dot\alpha}$). For instance, the symmetry of the
expression in Eq.~\ref{eq:rational-functions-1} would be denoted as
\texttt{({\color{BrickRed}{`12354'}}, {\color{blue}False})}.

\section{Ideals in spinor space}

Before constructing singular phase-space configurations, let us
consider how to make these special configurations unambiguous. To this
aim, we rely on the algebro-geometric concept of an ideal.

In \texttt{lips}, two classes are used to represent ideals:
\texttt{LipsIdeal} and \texttt{SpinorIdeal}. The former represents
covariant ideals in the ring of spinor components, while the latter
represents invariant ideals in the ring of spinor contractions
\cite{DeLaurentis:2022otd}. Both ideal types are subclasses of the
\texttt{Ideal} class of the interface \texttt{syngular}, through which
one can access many algorithms implemented in
\texttt{Singular}. Despite most applications of interest deal with
Lorentz invariants, it is generally convenient to work with spinor
components, as the ideals are generated by fewer polynomials.

To instantiate a \texttt{LipsIdeal} object one has to declare the
multiplicity of phase space, and a set of generating polynomials. For
instance, taking invariants which appear in
Eq.~\eqref{eq:rational-functions-1}, we can write \linebreak
\vspace{-3mm}
\begin{minted}[escapeinside=||, mathescape=true, linenos=False, numbersep=5pt, gobble=2, frame=lines, framesep=2mm, breaklines, breakautoindent=false, breakindent=-12.5pt, baselinestretch=0 ]{python}
  J = LipsIdeal(|\color{OliveGreen}{|5|}|, (|\color{OliveGreen}{|"<4|1+5|4]"|}|, |\color{OliveGreen}{|"<5|1+4|5]"|}|))
\end{minted}
\vspace{-1.5mm}
Momentum conservation is added by default. The method
\texttt{make\_analytical\_d} of the \texttt{Particle} class is used to
replace lower-index spinors with \texttt{sympy} symbols. Tensor
contractions are then computed as in numeric cases. The resulting
expressions are then passed to \texttt{Singular} for subsequent
manipulations. The default ring is a polynomial ring, while the
quotient ring by the momentum-conserving ideal can be accessed via the
method \texttt{to\_mom\_cons\_qring}. This modifies the ideal in
place. Translation to a \texttt{SpinorIdeal}, i.e.~to the Lorentz
invariant subring, can be computed with the method
\texttt{invariant\_slice}. This relies on an elimination theory
algorithm.

To understand the geometry of a variety associated to an ideal we need
to compute its primary decomposition. Prime ideals will then
correspond to irreducible varieties, i.e.~phase space configuration
where amplitudes have well-defined poles and zeros. To obtain the
prime ideals associated with a given ideal, such as the above defined
\texttt{J}, one can (try to) compute a primary decomposition via
\texttt{Singular}. This approach is in general insufficient to map out
the irreducible varieties, due to the large number of variables in the
underlying ring. However, we can use physical understanding to gain
insights into the primary decomposition. For instance, the ideal
\texttt{J} has five branches, but only one is non-trivial.
\begin{minted}[escapeinside=||, mathescape=true, linenos=False, numbersep=5pt, gobble=2, frame=lines, framesep=2mm, breaklines, breakautoindent=false, breakindent=-12.5pt, baselinestretch=0 ]{python}
  K = LipsIdeal(|\color{OliveGreen}{|5|}|, (|\color{OliveGreen}{|"<14>"|}|, |\color{OliveGreen}{|"<15>"|}|, |\color{OliveGreen}{|"<45>"|}|, |\color{OliveGreen}{|"[23]"|}| ))
  L = LipsIdeal(|\color{OliveGreen}{|5|}|, (|\color{OliveGreen}{|"<12>"|}|, |\color{OliveGreen}{|"<13>"|}|, |\color{OliveGreen}{|"<14>"|}|, |\color{OliveGreen}{|"<15>"|}|, |\color{OliveGreen}{|"<23>"|}|, |\color{OliveGreen}{|"<24>"|}|, |\color{OliveGreen}{|"<25>"|}|, |$\qquad\qquad\qquad\qquad$\color{OliveGreen}{|"<34>"|}|, |\color{OliveGreen}{|"<35>"|}|, |\color{OliveGreen}{|"<45>"|}|))
  M = LipsIdeal(|\color{OliveGreen}{|5|}|, (|\color{OliveGreen}{|"<4|1+5|4]"|}|, |\color{OliveGreen}{|"<5|1+4|5]"|}|, |$\qquad\qquad\qquad\qquad$\color{OliveGreen}{|"|1]<14><15>+|4]<14><45>-|5]<45><15>"|}|, |$\qquad\qquad\qquad\qquad$\color{OliveGreen}{|"|1>[14][15]+|4>[14][45]-|5>[45][15]"|}|))
\end{minted}
\vspace{-1.5mm}
The ideals \texttt{K} and \texttt{L} are essentially
triple-collinear configurations for legs 1, 4 and 5. Once these are
known, the ideal \texttt{M} can be easily obtained by computing ideal
quotients. To check the decomposition, we can compute an intersection
of ideals with the operator \texttt{\&}, which calls the command
\texttt{intersect} of \texttt{Singular}, and verify the equality, via
reduced Gr\"obner bases.
\begin{minted}[escapeinside=||, mathescape=true, linenos=False, numbersep=5pt, gobble=2, frame=lines, framesep=2mm, breaklines, breakautoindent=false, breakindent=-12.5pt, baselinestretch=0 ]{python}
  assert J == K & K("12345", True) & L & L("12345", True) & M
\end{minted}
\vspace{-1.5mm} Note the use of another magic method
(\texttt{\_\_call\_\_}) to compute the image of an ideal under a
symmetry of phase space, similarly to the \texttt{Particles.image}
method. By itself, this assertion is not sufficient to prove a primary
decomposition of the ideal $\texttt{J}$. One also has to prove that
the ideals \texttt{K}, \texttt{L}, \texttt{M} are prime. An efficient
way to do this is via a prime test implement in $\texttt{syngular}$,
which can be accessed via the method \texttt{test\_primality}. This
assumes equi-dimensional ideals.

\section{Singular varieties}

In \texttt{lips} a singular variety, irrespective of its dimension, is
represented by single, generic, phase-space point---i.e.~a
zero-dimensional variety---embedded in the multi-dimensional
variety. In this context, generic means that the result of evaluations
at the chosen phase-space point should have an absolute value
representative of evaluations at most points on the variety, possibly
barring special, higher-codimension embedded varieties. This is
guaranteed with high probability by picking the point at random, while
satisfying the constraints that define the variety.

Two facilities are provided for the generation of finely-tuned
phase-space points on specific varieties. First, in the submodule
\texttt{hardcoded\_limits} two methods, \texttt{\_set} and
\texttt{\_set\_pair}, are implemented. These methods efficiently
generate points on varieties of codimension one and two
respectively. However, they do not know about primary
decompositions. As such, in case a variety is multi-branched, a branch
will be chosen at random. Furthermore, since the constraints are
solved explicitly, only some configurations can be built this way. The
second option is to use the \texttt{algebraic\_geometry} submodule,
where the method \texttt{\_singular\_variety} is implemented. This
method is significantly more computationally intensive than the
former, as it relies on lexicographic Gr\"oebner bases, but it allows
to specify branches---i.e. irreducible sub-varieties.

For instance, the following code will generate a 3-digits
$2147483647$-adic phase-space point near the irreducible variety
associated to the prime ideal \texttt{M}.
\begin{minted}[escapeinside=||, mathescape=true, linenos=False, numbersep=5pt, gobble=2, frame=lines, framesep=2mm, breaklines, breakautoindent=false, breakindent=-12.5pt, baselinestretch=0 ]{python}
  oPsM = Particles(|\color{OliveGreen}{|5|}|, field=Field("padic", |\color{OliveGreen}{|2 ** 31 - 1|}|, |\color{OliveGreen}{|3|}|), seed=|\color{OliveGreen}{|0|}|)
  oPsM._singular_variety(("⟨4|1+5|4]", "⟨5|1+4|5]"), (|\color{OliveGreen}{|1|}|, |\color{OliveGreen}{|1|}|),
  |$\;\qquad\qquad\qquad\qquad\qquad$|generators=M.generators)
\end{minted}
\vspace{-2mm} The first argument specifies orthogonal directions to
the variety, the second the valuations of the invariants in the first
argument (in this case both proportional to the chosen prime), while
the \texttt{generators} keyword argument specifies the
branch. Asymmetric limits can also be constructed by providing unequal
valuations, see ref.~\cite[Appendix C]{Campbell:2022qpq}.

\section{Partial fraction decompositions}

Partial fraction decompositions play an important role in the
computation of scattering amplitudes, both in terms of final
expressions, as well as at intermediate stages, e.g.~for
integration-by-parts identities. Standard methods are based on
symbolic computations, including Gr\"obner bases and polynomial
reduction. For instance, see ref.~\cite{Heller:2021qkz}. We can use
the technology described in the previous sections to infer whether a
given partial fraction decomposition is valid, before determining the
analytic form of the numerator. Given denominator factors
$\mathcal{D}_1$ and $\mathcal{D}_2$,
\begin{itemize}
\item[1.] compute the primary decomposition for the ideal $\big\langle
  \mathcal{D}_1, \mathcal{D}_2 \big\rangle$;
\item[2.] generate a phase-space point near each branch of the variety
  $V(\big\langle \mathcal{D}_1, \mathcal{D}_2 \big\rangle)$;
\item[3.] numerically evaluate the function at these points.
\end{itemize}
If the numerator vanishes on all of them, then it belongs to (the
radical of) the associated ideal (Hilbert's Nullstellensatz). For
instance, given the expression of Eq.~\eqref{eq:rational-functions-1},
we can infer that the denominator factors $⟨4|1+5|4]$ and $⟨5|1+4|5]$
must be separable into different fractions because the numerator,
considered in least common denominator form, vanishes on all 5
branches. Further constraints from the degree of vanishing can also be
imposed via the Zariski--Nagata theorem \cite{DeLaurentis:2022otd}.

\vspace{-3mm}
\paragraph{Beyond partial fractions.}
Partial fraction decompositions deal purely with sets of denominator
factors, i.e.~the poles of the functions. Yet, even if no partial
fraction decomposition is possible, for instance when the denominator
is a single irreducible polynomial, the numerator may still have a
simple structure, generally in terms of an expanded set of
invariants. These new invariants can be systematically identified via
primary decompositions, and the same logic described to separate poles
in the denominators can also be used to identify factors of the
numerators.

\vspace{-2mm}
\section*{References}
{\tiny
\bibliography{lips}

\providecommand{\newblock}{}
\begin{thebibliography}{10}
\expandafter\ifx\csname url\endcsname\relax
  \def\url#1{{\tt #1}}\fi
\expandafter\ifx\csname urlprefix\endcsname\relax\def\urlprefix{URL }\fi
\providecommand{\eprint}[2][]{\url{#2}}

\bibitem{DGPS}
Decker W {\em et~al.\/} {\sc Singular} {4-2-1} -- {A} computer algebra system
  for polynomial computations

\bibitem{vonManteuffel:2014ixa}
von Manteuffel A and Schabinger R~M 2015 {\em Phys. Lett. B\/} {\bf 744}
  101--104 \href{https://arxiv.org/abs/1406.4513}{{\ttfamily 1406.4513}}

\bibitem{Peraro:2016wsq}
Peraro T 2016 {\em JHEP\/} {\bf 12} 030
  \href{https://arxiv.org/abs/1608.01902}{{\ttfamily 1608.01902}}

\bibitem{DeLaurentis:2019bjh}
De~Laurentis G and Ma\^\i{}tre D 2019 {\em JHEP\/} {\bf 07} 123
  \href{https://arxiv.org/abs/1904.04067}{{\ttfamily 1904.04067}}

\bibitem{DeLaurentis:2022otd}
De~Laurentis G and Page B 2022 {\em JHEP\/} {\bf 12} 140
  \href{https://arxiv.org/abs/2203.04269}{{\ttfamily 2203.04269}}

\bibitem{Abreu:2020cwb}
Abreu S, Page B, Pascual E and Sotnikov V 2021 {\em JHEP\/} {\bf 01} 078
  \href{https://arxiv.org/abs/2010.15834}{{\ttfamily 2010.15834}}

\bibitem{Chawdhry:2020for}
Chawdhry H~A, Czakon M, Mitov A and Poncelet R 2021 {\em JHEP\/} {\bf 06} 150
  \href{https://arxiv.org/abs/2012.13553}{{\ttfamily 2012.13553}}

\bibitem{Abreu:2023xxx}
Abreu S, De~Laurentis G, Ita H, Klinkert M, Page B and Sotnikov V
  \href{https://arxiv.org/abs/23xx.xxxxx}{{\ttfamily 23xx.xxxxx}}

\bibitem{binder}
Jupyter {\em et~al.\/} Binder 2.0 - reproducible, interactive, sharable
  environments for science at scale.

\bibitem{2020NumPy-Array}
Harris C~R {\em et~al.\/} 2020 Array programming with {NumPy}

\bibitem{mpmath}
Johansson F {\em et~al.\/} 2013 mpmath: a {P}ython library for
  arbitrary-precision floating-point arithmetic

\bibitem{10.7717/peerj-cs.103}
Meurer A {\em et~al.\/} 2017 Sympy: symbolic computing in python

\bibitem{Lenstra1982}
Lenstra H {\em et~al.\/} 1982 {\em Mathematische Annalen\/} {\bf 261} 515--534
  \urlprefix\url{http://eudml.org/doc/182903}

\bibitem{inproceedings}
Monagan M 2004 pp 243--249

\bibitem{Klappert:2019emp}
Klappert J and Lange F 2020 {\em Comput. Phys. Commun.\/} {\bf 247} 106951
  \href{https://arxiv.org/abs/1904.00009}{{\ttfamily 1904.00009}}

\bibitem{Hostetter_Galois_2020}
Hostetter M 2020 {Galois: A performant NumPy extension for Galois fields}
  \href{https://github.com/mhostetter/galois}{github.com/mhostetter/galois}

\bibitem{syngular}
De~Laurentis G 2021 syngular
  \href{https://github.com/GDeLaurentis/syngular}{github.com/GDeLaurentis/syngular}

\bibitem{DeLaurentis:2020xar}
De~Laurentis G {\em {Numerical techniques for analytical high-multiplicity
  scattering amplitudes}\/} Ph.D. thesis

\bibitem{Campbell:2022qpq}
Campbell J~M, De~Laurentis G and Ellis R~K 2022 {\em JHEP\/} {\bf 07} 096
  \href{https://arxiv.org/abs/2203.17170}{{\ttfamily 2203.17170}}

\bibitem{Heller:2021qkz}
Heller M and von Manteuffel A 2022 {\em Comput. Phys. Commun.\/} {\bf 271}
  108174 \href{https://arxiv.org/abs/2101.08283}{{\ttfamily 2101.08283}}

\end{thebibliography}
}

\end{document}